\documentclass{elsart}

\usepackage{natbib}

   \usepackage{graphicx}
  \usepackage{amssymb}

\def\teq#1{$\, #1\,$}                         
\def\apj{ApJ}
\def\apjsupp{ApJ Supp.}

\def\asr{Adv. Space Res.}                       
\def\grl{Geophys. Res. Lett.}                   
\def\jgr{J. Geophys. Res.}
\def\mnras{{M.N.R.A.S.}}
\def\ssr{Space Sci. Rev.}                       
\def\reference{\par \noindent \hangafter=1 \hangindent=0.7 true cm}
          \font\sixrm=cmr6

\def\machson{{M}_{\hbox{\sixrm S}}}
\def\machalf{{M}_{\hbox{\sixrm A}}}
\def\machtot{{M}_{\hbox{\sixrm T}}}
\def\dover#1#2{\hbox{${{\displaystyle#1 \vphantom{(} }\over{
   \displaystyle #2 \vphantom{(} }}$}}

\begin{document}
\newcommand{\vol}[2]{$\,$\rm #1\rm , #2.}                 
\newcommand{\figureoutpdf}[5]{\centerline{}
   \centerline{\hspace{#3in} \includegraphics[width=#2truein]{#1}}
   \vspace{#4truein} \caption{#5} \centerline{} }
\begin{frontmatter}

\title{Modeling Accelerated Pick-up Ion Distributions at an Interplanetary Shock}

\author{Errol J. Summerlin \& Matthew G. Baring}

\address{Department of Physics and Astronomy, MS-108,
Rice University, P. O. Box 1892, Houston, TX 77251-1892, USA\\
{\tt Emails: xerex@rice.edu, baring@rice.edu}}

\begin{abstract}
The acceleration of interstellar pick-up ions as well as solar wind
species has been observed at a multitude of interplanetary (IP) shocks
by different spacecraft.  The efficiency of injection of the pick-up ion
component differs from that of the solar wind, and is expected to be
strongly enhanced at highly oblique and quasi-perpendicular shock
events, in accord with inferences from {\it in situ} observations.  This
paper explores theoretical modeling of the phase space distributions of
accelerated ions obtained by the Ulysses mission for the Day 292, 1991
shock associated with a corotating interaction region, encountered
before Ulysses' fly-by of Jupiter.  A Monte Carlo simulation is used to
model the acceleration process, adapting a technique that has been
successfully tested on earlier IP shocks possessing minimal pick-up ion
presence.  Phase space distributions from the simulation technique for
various low mass ions are compared with SWICS and HI-SCALE data to
deduce values of a ``turbulence parameter'' that controls the efficiency
of injection, and the degree of cross-field diffusion.  Acceptable fits
are obtained for the \teq{H^+} and \teq{He^+} populations using standard
prescriptions for the pick-up ion distribution; \teq{He^{++}} spectral
data was only fit well for scenarios very close to the Bohm diffusion
limit.  It is also found that the simulation successfully accounts for
the observation of energetic protons farther upstream of the forward
shock than lower energy pick-up protons, using the same turbulence
parameter that is required to achieve reasonable spectral fits.
\end{abstract}

\begin{keyword}
Interplanetary shocks \sep Co-rotating interaction regions \sep Ulysses
mission \sep Diffusive shock acceleration \sep Shock turbulence
\end{keyword}

\end{frontmatter}
\setlength{\parindent}{.25in}

\newpage 

\section{Introduction}
\label{sec:Introduction}

Particle acceleration at collisionless shocks is believed to be a common
phenomenon in space plasmas in a diversity of environments, ranging from
the inner heliosphere to the central regions of distant galaxies.  In
the heliosphere, evidence to support the belief that such a mechanism
can efficiently produce non-thermal particles includes direct
measurements of accelerated populations in various energy ranges at the
Earth's bow shock (e.g. Scholer et al. 1980; M\"obius et al. 1987;
Gosling et al. 1989) and interplanetary shocks (for the pre-Ulysses era
see, for example, Sarris and Van Allen 1974; Gosling et al. 1981;
Decker, Pesses and Krimigis, 1981; Tan et al. 1988).  The motivations
for developing theories of shock acceleration are therefore obvious, and
a variety of approaches have emerged.  One possible means for the
generation of non-thermal particles is the Fermi mechanism, often called
diffusive shock acceleration; it is this process that is the focus of
this paper.

There are various approaches to modelling diffusive shock acceleration.
Among these are  convection-diffusion differential equation approaches
(e.g. Kang and Jones 1995), and the kinematic Monte Carlo technique of
Ellison and Jones (e.g., see Jones and Ellison, 1991, and references
therein), which describes the injection and acceleration of particles
from thermal energies; both of these can  address spectral and
hydrodynamic properties.  This latter simulational approach is the
central tool for the analysis here, and is ideally suited to the
interpretation of the modest time-resolution shock data acquired by
Ulysses.  Hybrid and full plasma codes (e.g. Quest 1988; Burgess 1989,
Winske et al. 1990; Trattner and Scholer 1991; Giacalone, Burgess and
Schwartz 1992; Liewer, Goldstein, and Omidi 1993; Kucharek and Scholer
1995) provide contrasting probes of shock environs, with an emphasis
primarily on plasma structure and wave properties in the environs of
shocks.

Comparisons of predictions from theoretical models with observed
phase-space distribution data are informative probes.  The first
detailed theory/data comparison along these lines was performed by
Ellison, M\"obius and Paschmann (1990) in the case of the quasi-parallel
portion of the Earth's bow shock, comparing predictions of the Monte
Carlo method with particle distributions of protons, \teq{He^{++}} and a $C$,
$N$ and $O$ ion mix obtained by the AMPTE experiment.  The agreement 
between model predictions and data was impressive.  Ellison, M\"obius and
Paschmann (1990) concluded that a successful fit was possible only in
the non-linear acceleration regime, when the dynamic effects of the
accelerated particles are crucial to the determination of the shock
structure. Since this pioneering work, successful comparisons of other
theoretical techniques with data from the Earth's bow shock have been
performed.  These include the hybrid plasma simulations of Trattner and
Scholer (1991) and Giacalone et al. (1993), and solutions to the
convection-diffusion equation (Kang and Jones 1995), both of which have
yielded good agreement with both the bow shock data and the Monte Carlo
technique (Ellison et al. 1993).

Such a comparison between theory and experiment was extended to the
domain of interplanetary shocks in the work of Baring, Ogilvie, Ellison
and Forsyth (1997, hereafter BOEF97; see also Kang \& Jones 1997, for
application of their convection-diffusion equation technique).  In this
development, impressive agreement was found between the Monte Carlo
predictions and spectral data obtained by the Solar Wind Ion Composition
Spectrometer (SWICS) aboard Ulysses, in the case of two shocks observed
early in the Ulysses mission.  Such agreement was possible only with the
assumption of strong particle scattering (i.e. near the Bohm diffusion
limit) in the highly oblique candidate shocks.  For a third shock,
detected a month later, the comparison failed, with significant
differences arising in the 500-800 km/sec range of the phase space
distribution.  BOEF97 attributed this discrepancy to the omission of
pick-up ions from the model: such an extra component would be expected
to provide a substantial contribution to the accelerated population in
this particular event.

This paper addresses the role of pick-up ions in such shocks via
modeling the accelerated population for the specific interplanetary
shock observed on day 292 of 1991, detected by the SWICS and HI-SCALE
instruments aboard the Ulysses spacecraft at around 4.5 AU, as reported
in Gloeckler et al. (1994).  Phase space distributions from the
simulations are compared with SWICS and HI-SCALE data, yielding
acceptable fits for the \teq{H^+} and \teq{He^{++}} populations, using
standard prescriptions for the injected pick-up ion distribution, by
adjusting a single turbulence parameter \teq{\eta}.  Using this same
\teq{\eta}, the simulation results also successfully account for the
observation of energetic protons farther upstream of the forward shock
than lower energy pick-up protons, since a rigidity-dependent diffusion
is used in the models.

\section{The Monte Carlo Simulation Technique}
 \label{sec:mc}

The Monte Carlo simulation technique used here has been invoked in
previous applications to shocks in the heliosphere, as discussed above,
and is extensively described in a number of papers (Ellison, Jones, \&
Eichler 1981; Ellison, Jones \& Reynolds 1990; Jones \& Ellison 1991;
Baring, Ellison \& Jones 1994; Ellison, Baring, \& Jones 1996).  It
follows closely Bell's (1978) test particle approach to diffusive
acceleration. Particles are injected upstream and allowed to convect
into the shock, meanwhile diffusing in space so as to effect multiple
shock crossings, and thereby gain energy through the shock drift and
Fermi processes.  For the interplanetary shock that is the focus of this
paper, throughout  \teq{u_1} and \teq{u_2} shall denote the upstream and
downstream flow speed in the shock rest frame, respectively.

The particles gyrate in a laminar electromagnetic field.  Their
trajectories are obtained by solving a fully relativistic Lorentz force
equation in the shock rest frame, in which there is, in general, a {\bf
u $\times$ B} electric field in addition to the magnetic field. The
effects of magnetic turbulence are modeled by scattering these ions in
the rest frame of the local fluid flow.  These collisions, effectively
mimicking diffusive transport instigated by Alfv\'en waves, are assumed
to be elastic in the local fluid frame, an assumption that is valid so
long as the flow speed far exceeds the Alfv\'{e}n velocity.  Otherwise,
the scattering centers are not anchored in the fluid frame, and the
collisions are inelastic, thereby introducing significant second-order
Fermi (stochastic) acceleration.  For interplanetary shocks in general,
and in particular for the Day 292, 1991 shock considered here, the
elastic scattering assumption is somewhat violated due
to a low shock Alf\'{v}enic Mach number.  Yet the stochastic
acceleration contribution is expected to be relatively small, perhaps at
most of the order of 25\%, so its inclusion is deferred to later work.

The simulation can routinely model either large-angle or small-angle
scattering. In this paper, large-angle scattering is employed, motivated
by previous observations (e.g., Hoppe et al. 1981; Balogh et al. 1993)
indicating very turbulent fields in IP shocks.  Between scatterings,
particles are allowed to travel a distance in their local fluid frame
that is exponentially distributed about the scattering mean free path
\teq{\lambda}, with: 
\begin{equation}
	\lambda \; =\; \lambda _{0} \left( \frac{r_{g}}{r_{g1}} \right)^{\alpha}
	 \; \propto\; p^{\alpha}\quad .
 \label{eq:mfp}
\end{equation}
Here \teq{r_{g}= pc/(qB)} is the gyroradius of an ion of momentum
\teq{p=mv}, mass \teq{m}, and charge \teq{q}.   Also 
\teq{r_{g1}=mu_{1.x}c/(qB)} is the gyroradius of an ion with a speed 
\teq{v} equal to the far upstream flow speed normal
to the shock plane, \teq{u_{1.x}}, where \teq{x} denotes the direction normal
to the shock plane.  The mean free path scale \teq{\lambda_0} is set
proportional to \teq{r_{g1}} with constant of proportionality \teq{\eta}
such that \teq{\lambda_0 = \eta r_{g1}}. Following previous Monte Carlo
work, for simplicity we set \teq{\alpha =1}, a specialization that is
appropriate for interplanetary plasma shocks (e.g., see Ellison et al.
1990; Mason, Gloeckler \& Hovestadt 1983; Giacalone, Burgess \& Schwartz
1992 for discussions on the micro-physical expectations for 
\teq{\alpha}), yet is easily generalizable in the simulation.  Since
\teq{\lambda\geq r_g} for physically meaningful diffusion resulting from
gyro-resonant wave-particle interactions, the \teq{\alpha =1} case is
also motivated on fundamental grounds.

Cross-field diffusion emerges naturally from the simulation, since at
every scattering, the direction of the particle's momentum vector is
randomized in the local fluid frame, with the resulting effect that the
gyrocenter of a particle is shifted randomly by a distance of the order
of one gyroradius in the plane  orthogonal to the local field. 
Transport perpendicular to the field is then governed by a kinetic
theory description, so that the ratio of the spatial diffusion
coefficients parallel (\teq{\kappa_\parallel =\lambda v/3}) and
perpendicular  (\teq{\kappa_\perp}) to the mean magnetic field is given
by \teq{\kappa_\perp /\kappa_\parallel = 1/(1+\eta^2)} (see Forman,
Jokipii \& Owens 1974; Ellison, Baring \& Jones 1995, for detailed
expositions). Hence, \teq{\eta} couples directly to the amount of
cross-field diffusion, and is a measure not only of the frequency of
collisions between particles and waves, but also of the level of
turbulence present in the system, i.e. is an indicator of \teq{\langle
\delta B/B\rangle}. Note that \teq{\eta =1} is the Bohm diffusion limit
of quasi-isotropic diffusion, presumably corresponding to \teq{\langle
\delta B/B\rangle\sim 1}. As will become evident in the
Section~\ref{sec:model}, by virtue of its connection to cross-field
diffusion, \teq{\eta} plays an important role in determining the
injection efficiency of low energy particles.  We note that other
cross-field transport effects such as that incurred by so-called
field-line wandering (e.g. see Giacalone \& Jokipii 1999) can be
incorporated parametrically in the Monte Carlo technique, and will be
addressed in future work.

While \teq{\eta} and \teq{\alpha} serve as parameters here, in principal
they can be calculated for a given wave field.  However, in practice,
precision is limited by {\it in situ} magnetometer data, and exact
determination of 3-dimensional particle diffusion properties from a
1-dimensional field data stream is impossible.  Yet insights can be
gained by using established theoretical formalism, such as random
fluctuation theory (e.g. Jokipii \& Coleman 1968), or equivalently
quasi-linear diffusion theory (see, e.g., Kulsrud \& Pierce 1969; Jones,
Birmingham, \& Kaiser 1978), to estimate components of the diffusion
tensor along and orthogonal to the mean field.  These components can be
expressed in terms of integrals of components of the power spectral
tensor for field fluctuations.  Such slab models for turbulence can be
extended to consider higher-dimensional field fluctuations (e.g. see
Bieber et al. 1994). Guidance for the behavior and values of
\teq{\alpha} and \teq{\eta} can be gleaned from such investigations, and
an analysis along these lines to facilitate the interpretation of
Ulysses IP shock spectral data will be the subject of future work.

The simulation technique makes no distinction between accelerated
particles and thermal ones. Therefore the injection efficiency is
determined solely by the a particle's ability to diffuse back upstream
after it has encountered the shock for the first time. In the case of
the Earth's bow shock (Ellison et al. 1990), this permits the
development of a nonlinear model that includes the effects of the
accelerated particles on the dynamics of the shock. However, in the
relatively weak IP shock considered here, a steep power-law is present,
and the poor injection efficiency reduces the number density of
energetic particles, eliminating nonlinear hydrodynamic effects.
Upstream plasma quantities are input from observational data, as
discussed below, and downstream quantities are determined using the full
MHD Rankine-Hugoniot relations. The simulation is capable of producing
energy spectra/fluxes, and hence phase space distributions, at any
location upstream or downstream of the shock and in any reference frame
including that of a spacecraft. This makes the simulation ideal for
comparison with observational data.

\section{Modeling the Ulysses Event of Day 292, 1991}
 \label{sec:model}

Our case study focuses on the forward shock of a CIR encountered by
Ulysses on Day 292 of 1991, for which downstream particle distributions
were obtained for comparison with data published in Gloeckler et al.
(1994). To determine parameters for this shock that are appropriate
input for the Monte Carlo simulation,  the observational results
detailed in Gloeckler et al. (1994) are used along with the data
compilations of Balogh et al. (1995), Gonz\'alez-Esparza et al. (1996),
and Hoang et al. (1995).  These sources indicate that the sonic Mach
number \teq{M_{\rm s}} of the shock was 2.53, with an unstated
uncertainty. This controls the temperature of the thermal solar wind
protons, but does receive contributions from electrons and alpha
particles.  Other parameters include the angle \teq{\theta_{Bn1} =
50^{\circ}\pm 11^{\circ}} the upstream magnetic field makes with the
shock normal, \teq{B_1=0.8}nT, and magnetic compression ratio
\teq{B_{2}/B_{1} = 2.50 \pm 0.13}. Here \teq{B_{1}} and \teq{B_{2}} are
the {\it total} magnetic field magnitudes upstream and downstream of the
shock, respectively. In addition, the normalization of solar wind
distributions was established using \teq{n_p=2.0}cm$^{-3}$ as the solar
wind proton density, and \teq{n_{He}=4.0\times 10^{-2}}cm$^{-3}$ as that
for solar wind \teq{He^{++}} (within the uncertainties of values listed
Table~1 in Gloeckler et al. 1994).

Although fluctuations in the plasma data preclude precise determination
of the velocity compression ratio of the shock, Gloeckler et al.
inferred a value of \teq{r=u_1/u_2=2.4 \pm 0.3}. This inference is
somewhat inconsistent with the observed \teq{v^{-5.75}} power-law
behavior of the phase space density at high energies, which suggests a
compression ratio nearer 2. The choice of compression ratio is
controlled by the shock Mach number \teq{\machtot =\machson\machalf/
\sqrt{\machson^2+\machalf^2}}, receiving contributions from sonic and
Alfv\'enic Mach numbers, \teq{\machson} and \teq{\machalf} respectively,
which can be written in the form (e.g. Baring et al. 1997):
\begin{equation}
   \machson\;\approx\; 8.52\,\dover{u_{100}}{\sqrt{T_{p4}+T_{e4}}}\quad ,
   \quad \machalf\;\approx\; 4.58\,\dover{u_{100}}{B_{-5}}\,\sqrt{n_p}\quad ,
  \label{eq:machnos}
\end{equation}
where \teq{u_{100}} is the shock speed \teq{u_{1,x}} in units of 100 km/sec,
\teq{T_{p4}} and \teq{T_{e4}} are the proton and electron temperatures
in units of \teq{10^4}K, \teq{B_{-5}} is the upstream field strength in
units of \teq{10^{-5}}Gauss, and \teq{n_p} is the proton density in
\teq{\hbox{cm}^{-3}}.  Using upstream values of \teq{T_{e4}=2.59} and
\teq{T_{p4}=1.67}, as tabulated in Hoang et al. (1995), we find that for
\teq{u_{1,x}=55}km/sec, \teq{\machson\approx 2.5}. Adding in the magnetic
field gives  \teq{\machtot\approx 2.1}.  To yield a compression ratio
\teq{r=(\gamma+1)/[\gamma -1+2/\machtot^2] =2.1} for adiabatic index
\teq{\gamma=5/3}, one must have a slightly lower value of \teq{\machtot
\approx 1.85} (i.e. perhaps higher field) .  In this paper, we adopted a
value of \teq{r=2.1}, which lies within the margin of error of Gloeckler
et al.'s inferred value, to generate power-law slopes commensurate with
the data.  Due to significant observational uncertainties, ranges of
parameters are permissible.

In the simulation runs, the temperature \teq{T_\alpha} of the
\teq{\alpha} particles was chosen to be four times that of the solar
wind protons.  No guidance for this assumed upstream temperature was
provided in Gloeckler et al. (1994), nor in the aforementioned Ulysses
data compilation papers.  However, in the IP shock theory/data
comparison of Baring et al. (1997), for the one shock with good
measurements of the \teq{He^{++}} distribution, namely the 91097 event,
the upstream \teq{\alpha} particle thermal velocity was comparable to
(actually slightly greater) than that of the solar wind protons,
implying \teq{T_\alpha \approx 4T_p}.  Adopting the same temperature
ratio \teq{T_\alpha /T_p} in this work was motivated by this precedent,
and the expectation that the adiabatic expansion of the solar wind from
2.7AU to 4.5AU should roughly preserve the ratios of the temperatures
per nucleon for these two ions.  Note that the correction to the sonic
Mach number in Eq.~(\ref{eq:machnos}) incurred by solar wind helium is
then approximately 4\%.

To facilitate transformation of simulation results measured in the shock
rest frame to the approximate spacecraft frame, so as to compare
directly with Gloeckler et al.'s published distributions, a value for
the solar wind plasma speed of \teq{v_{sw,2}=393 \pm 12} km/s was
assumed {\it after} the shock encounter.  This identifies the downstream
fluid frame.  No values for the shock speed were listed in Gloeckler et
al. (1994), as is common practice in other expositions (e.g. see Burton
et al. 1992; Baring et al. 1997).  Hence selection of a value for
\teq{u_{1,x}} consistent with other data became necessary. From the
long-term solar wind radial velocity traces in Figure~2 of
Gonz\'alez-Esparza et al. (1996) we can set a lower bound on the solar
wind speed of \teq{v_{sw,1}\gtrsim 330} km/s prior to the shock passage,
i.e. upstream.   Assuming collinear velocities, i.e. that the solar wind
is not deflected by the shock on large scales, we have the speed
relations \teq{u_{1,x}=v_{sh}-v_{sw,1}} and \teq{u_{2,x}=v_{sh}-v_{sw,2}} in
addition to \teq{r=u_{1,x}/u_{2,x}}, where \teq{v_{sh}} is the unknown shock
speed in the spacecraft frame. These solve to yield \teq{u_{1,x} \approx
55} km/s and \teq{u_{2,x} \approx 28} km/s for \teq{r=2.1} for
\teq{v_{sw,1}\sim 360} km/s; higher values of \teq{u_{1,x}} and \teq{u_{2,x}}
would be realized for smaller choices of \teq{v_{sw,1}}.

The other input for the Monte Carlo simulation is that for the incoming
pick-up ion distributions.  Gloeckler et al. (1994) used a simplified
form for these distributions; here we adopt the slightly more developed
expression used in Ellison, Jones \& Baring (1999) that is modeled on
the seminal work of Vasyliunas \& Siscoe (1976).  This form incorporates
the gravitational focusing of interstellar neutrals, the physics of
their ionization as a function of distance from the sun (approximately
at solar minimum), and adiabatic losses incurred during propagation away
from the sun, and is similar in conception to pick-up ion distributions
used in le Roux, Potgieter \& Ptuskin (1996). Our pick-up ion model
therefore provides both the detailed shape and normalization of these
superthermal distributions, which are distinctly different for \teq{H^+}
and \teq{He^+}, largely due to the patently different ionization and
charge exchange rates for their corresponding interstellar neutrals.

\begin{figure}
\figureoutpdf{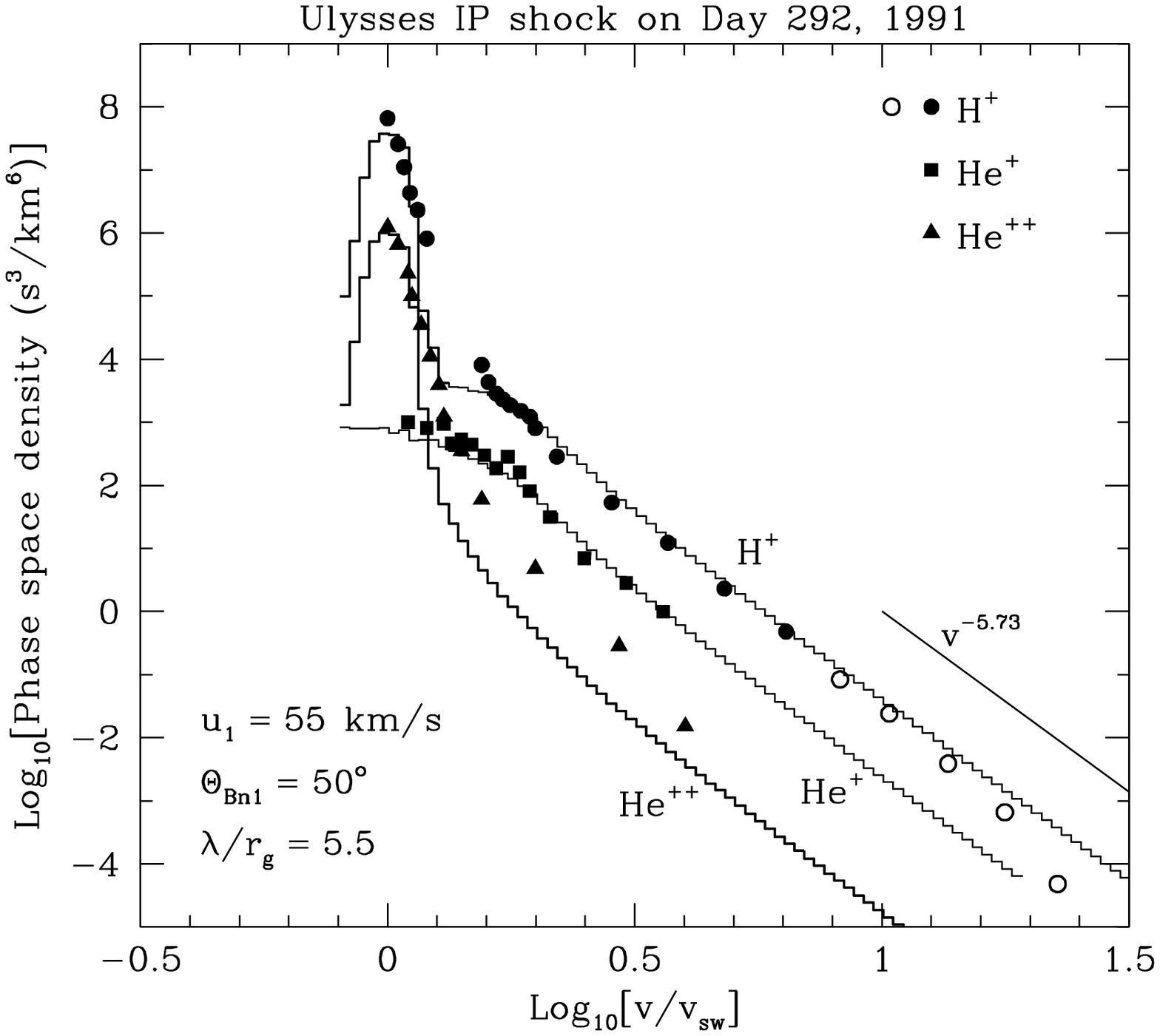}{5.5}{0.0}{-0.2}{
Phase space velocity distribution functions for data collected by the
Ulysses mission for the shock on Day 292 of 1991, specifically for the
interval 1991.292.0400-293.0400.   The velocity axis is the ratio of the
ion speed \teq{v}, as measured in the spacecraft frame, to the solar
wind speed.  The data are for \teq{H^+} (filled circles for SWICS data;
open circles for HI-SCALE points) solar wind and pickup ions, \teq{He^+}
(filled triangles) pickup ions, and \teq{He^{++}} (filled squares),
mostly solar wind ions, and are taken from Gloeckler et al. (1994).  The
histograms are the corresponding Monte Carlo models of acceleration of
these species (heavy weight corresponds to solar wind ions) for
\teq{u_1=55}km/sec, using the optimal choice of plasma shock parameters
from Gloeckler et al. (1994) and sources indicated in the text.  The
model assumed \teq{\eta =\lambda /r_g=5.5} and a shock of compression ratio
\teq{r=2.1}, corresponding to diffusive acceleration power-laws of index
\teq{-5.73}, is indicated.
}
 \label{fig:distrib}
\end{figure}

Downstream distributions for thermal and accelerated ions from the Monte Carlo
simulation are compared with SWICS and HI-SCALE measurements taken in
the frame of the spacecraft on the downstream side of the Day 292, 1991
shock in Fig.~\ref{fig:distrib}.  The Ulysses data are those exhibited
in Fig.~1 of Gloeckler et al. (1994), and the simulation data are
transformed to the spacecraft frame as described above.  The solar wind
and pick-up ion parameters and abundances are fairly tightly specified,
so that the theoretical model has one largely free parameter, the ratio
of the particle mean free path to its gyroradius, \teq{\eta =\lambda
/r_g}.  As noted in Ellison, Baring \& Jones (1995) and Baring et al.
(1997), the efficiency of acceleration of thermal ions in oblique
shocks, i.e. the normalization of the non-thermal power-law, is
sensitive to the choice of \teq{\eta}, so this parameter was adjusted to
obtain a reasonable ``fit'' to the Ulysses data. Ellison,
Jones \& Baring (1999) note that for their application to the solar wind
termination shock, the efficiency is not as sensitive to \teq{\eta} when
superthermal pick-up ions enter the problem, a property that we also 
replicate here.  Here, for \teq{\eta\gtrsim 3}, the accelerated
pick-up ion phase space density is orders of magnitude above that of 
the thermal ions, for both \teq{H} and \teq{He}.

In Fig.~\ref{fig:distrib}, the theory/data comparison is overall not
quite as good as the ones in Baring et al. (1997) for Ulysses IP shock
data at around 2--3 AU, where pick-up ions are less of a factor.  Yet
the fits here do model the accelerated pick-up ions very well, for
\teq{\eta = 5.5}, a value that is slightly higher than those
inferred in the fits of Baring et al. (1997), but is still consistent
with a moderate level of field turbulence (i.e. here
\teq{\kappa_\perp/\kappa_\parallel \approx 0.03}). Note that since the
inferred value of \teq{\eta =5.5} for \teq{H^+} and \teq{He^+} data is
most sensitive to the assumed shock obliquity \teq{\theta_{Bn1}} (among
other input parameters), the observational uncertainty in
\teq{\theta_{Bn1}} maps over to an uncertainty of around \teq{\pm 1.5}
in the value of \teq{\eta}.

There are clear differences between the simulation results and the
observations.  The \teq{He^{++}} distribution of thermal solar wind ions
appears slightly narrower than the published observations.  More notable
though is the fact that the accelerated thermal \teq{He^{++}} ions are
injected somewhat less efficiently in the simulation than in the
observations, an inefficiency characteristic of highly oblique shocks
that is present also for the {\it solar wind} protons, though not
explicitly displayed in Fig.~\ref{fig:distrib}.  The efficiency of
acceleration of thermal ions could be increased via several means: (i)
by lowering the shock obliquity angle \teq{\theta_{Bn1}}, for which
there is a large observational uncertainty; (ii) by decreasing
\teq{\eta}, corresponding to increased turbulence, without altering the
pick-up ion acceleration efficiencies substantially, and (iii) 
increasing the temperature of the thermal ions somewhat, though this
would reduce the compression ratio and accordingly steepen the
non-thermal continuum. However, such parameter adjustments only incur
small changes to the widths of the thermal distributions. Trial runs
indicate that an agreeable fit can be obtained to the entire
\teq{\alpha} particle distribution simply by reducing \teq{\eta} to
\teq{\eta \sim 1}, without changing other parameters.  It is not clear
what plasma characteristic would establish a species-dependent
\teq{\eta}, a circumstance that contrasts the inference of \teq{\eta
(H^+)\approx\eta (He^{++})} in the 91097 event  analyzed by Baring et
al. (1997).

\begin{figure}
\figureoutpdf{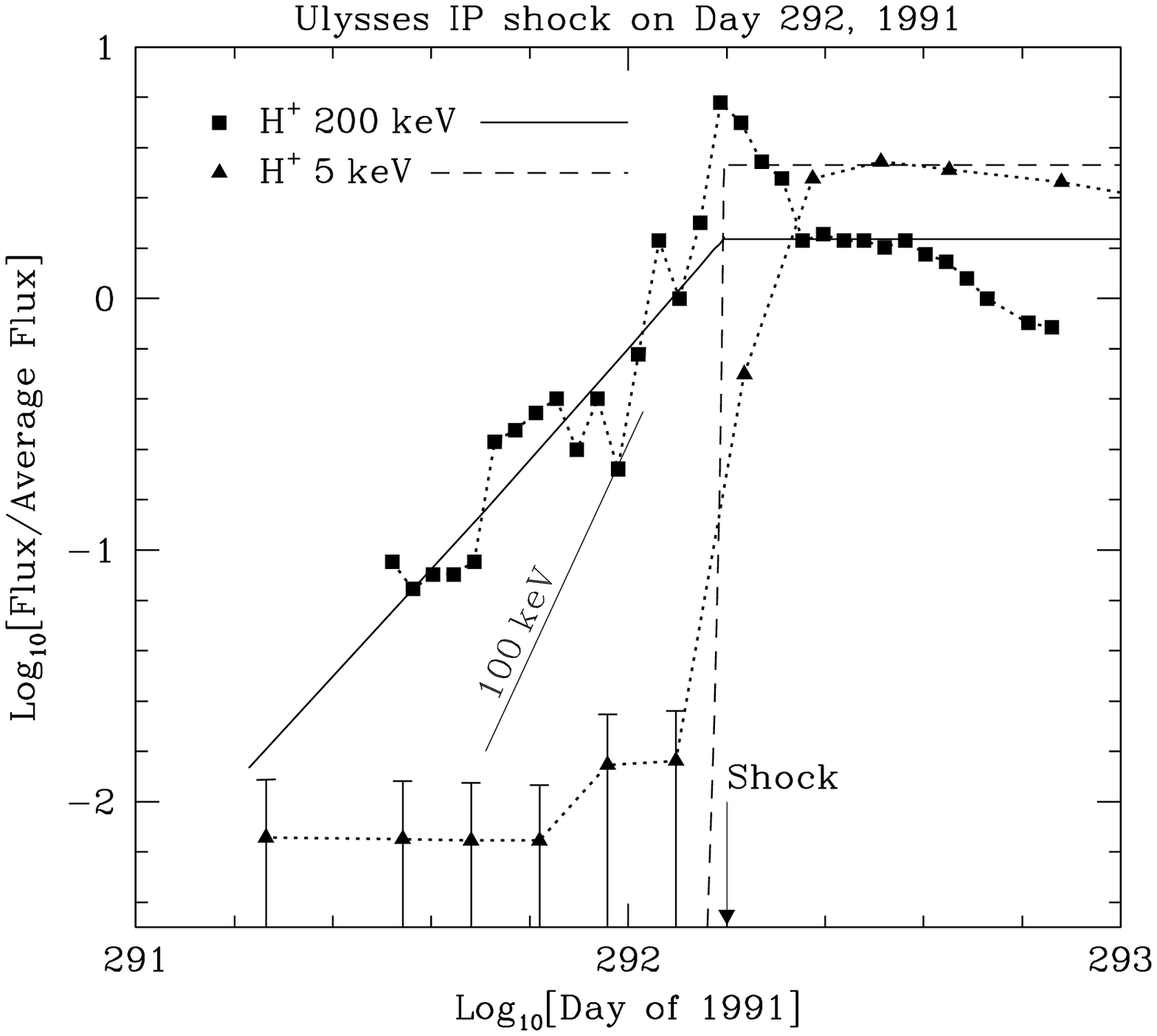}{5.5}{0.0}{-0.2}{
The flux variations of accelerated pick-up ion populations as a function
of time near the shock of Day 292, 1991.  The data for 5 keV and 200 keV
pick-up \teq{H^+} are depicted by filled triangles and squares,
respectively, and are taken from Gloeckler et al. (1994). The Monte
Carlo model generated fluxes at different distances normal to the shock,
and were converted to spacecraft times by incorporating solar wind
convection.  The 5 keV and 200 keV pick-up \teq{H^+} traces are
displayed as dashed and solid curves, respectively, and exhibit an
exponential decline upstream of the shock that is characteristic of
diffusive shock acceleration.  The model normalization protocol is
discussed in the text.  The lightweight line labelled ``100 keV'' is an
approximate indication of a model prediction for the expected flux
variation upstream for 100 keV protons.
}
\end{figure}

An instructive diagnostic on the acceleration model is to probe the
spatial scale of diffusion.  This is performed by examining upstream
distributions of high energy particles. In their 1994 paper, Gloeckler
et al. discussed an energy-dependent rise in fluxes of non-thermal
particles {\it prior} to the shock crossing. This was cited as
indicating the existence of a pre-acceleration mechanism.  The Monte
Carlo simulation was run with flux measurement planes placed upstream of
the shock at different distances, as well as downstream. This enabled
determination of the number fluxes of a particular energy ion at
different positions upstream of the shock.  This was converted to
observer's times using the approximate speed of the Ulysses spacecraft
in the solar wind. Results for two different \teq{H^+} ion energies are
displayed in Fig.~2, together with corresponding data from Fig.~3 of
Gloeckler et al. (1994) for identical energy windows.  \teq{H^+} was
chosen by Gloeckler et al. since good spatial resolution at 5 keV was
possible.   Observe that the y-axis in Fig.~2 represents ratios of flux
to time-averaged flux at the respective energies, so that the normalization
does not explicitly exhibit the property that the absolute flux at
200 keV is always much less than that at 5 keV when downstream of the shock.
Note also that the Ulysses data normalization was established by
averaging over 3 days of accumulated data, whereas the model 
normalization was adjusted to match observed fluxes around 1/2 day 
downstream of the shock.

The essential result of this comparison is that the spatial scale of the
exponential decline of ions upstream of the shock is more or less
identical to that of the model, for our choice of \teq{\eta=5.5}.
Theoretically, high energy particles with a mean free path according to
Eq.~(\ref{eq:mfp}) establish an exponential dilution in space/time due
to random scattering of the particles seeding upstream leakage against a
convective flow. From standard diffusion-convection theory, the spatial
scale for this dilution is \teq{(\kappa_\parallel \cos^2\theta_{Bn1} +
\kappa_\perp\sin^2\theta_{Bn1})/u_{1.x}} for a shock of obliquity
\teq{\theta_{Bn1}}, where \teq{\kappa_\parallel =\lambda v/3} is the
component of the spatial diffusion coefficient along the mean field,
\teq{\kappa_\perp =\kappa_\parallel /(1+\eta^2)}, and \teq{\lambda} is
prescribed by Eq.~(\ref{eq:mfp}).  Hence, this upstream dilution scale
is proportional to the proton's energy. For the 200 keV ions, the
simulation results are clearly well correlated with the data prior to
the shock, modulo plasma fluctuations.  On the other hand, for the lower
energy 5keV ions, the exponential decay has a very short time scale, a
factor of 40 smaller than for the 200 keV ions, and drops to background
levels very quickly.  So, although the simulation results are consistent
with the observed results, it is impossible to draw more definitive
conclusions without an improvement in data time resolution. The 200 keV
ions, however, have much longer mean free paths and travel much farther
upstream, and are well correlated.

\section{Conclusions}
\label{sec:conclusion}

The comparison of phase space distributions from the Monte Carlo
simulation of diffusive shock acceleration with those observed by SWICS
and HI-SCALE in the Day 292, 1991 event reveals a good deal of
consistency between theory and experiment for both the proton and
\teq{He^+} spectra above speeds around 600 km/sec.  At these speeds, the
injection of pick-up protons dominates that of solar wind protons, and
the acceleration of solar wind \teq{He^{++}} is inefficient relative to
that of pick-up \teq{He^+}. The characteristic of the shock most
relevant for these high energy components is its compression ratio
\teq{r}: to match the observed steep spectra, \teq{r\sim 2.1} is
required, on the low end of the range quoted by Gloeckler et al. (1994).

The normalization of the energetic ion power-laws is best fit with a
model ``turbulence'' parameter \teq{\eta = 5.5\pm 1.5}, where \teq{\lambda =\eta r_g},
corresponding to a ratio \teq{\kappa_\perp/\kappa_\parallel \approx 0.03}
of components of the spatial diffusion tensor.  For
such values, cross-field diffusion is insufficient for the Monte Carlo
model to account for observed accelerated \teq{He^{++}}, which
requires \teq{\eta\sim 1} so that the Bohm limit 
\teq{\kappa_\perp\sim\kappa_\parallel} is approximately realized.  
While the high energy protons and \teq{He^+} are modeled fairly well,
there are significant discrepancies in the high velocity
wings of the thermal protons.  These deviations are slightly 
more marked than those found by Baring et al. (1997) in IP shocks 
without significant pick-up ion components.  Clearly
some non-diffusive element of heating may be present in
the plasma shock that is not incorporated in the simulation.  To explore such
possibilities, it is planned to incorporate cross-shock and shock layer
charge separation potentials in the Monte Carlo code in future extensions,
to effect a more accurate modeling of the shock structure.  It must be noted,
though, that uncertainties in the shock parameters significantly impact
inferences of discrepancies between theory and experiment.

The flux increases of energetic protons seen upstream of the shock are
quite consistent with the expected upstream leakage associated with
diffusive shock acceleration.  The value of \teq{\eta =5.5} inferred from
the spectral fit scales the upstream diffusive lengthscale, and the
accompanying exponential decline in predicted flux is commensurate with
the Ulysses data presented in Gloeckler et al. (1994).  Hence, the
observed upstream flux precursor is not clear evidence of a
pre-acceleration mechanism, as claimed by Gloeckler et al., though it is
quite possible that some pre-acceleration mechanism may be acting.
A most enticing feature of this work is that a single 
model parameter can couple injection efficiency to the spatial scales 
of ions upstream of the shock.  Whether or not this identifies diffusion as
the dominant operating mechanism remains to be seen, but it is clear
that processes acting on the scales of a few gyroradii control both injection
and upstream transport in this interplanetary shock.

\section{References}

\vskip -5pt
\setlength{\parskip}{.00in}

\reference                                                      
Balogh, A., Forsyth, R. J., Ahuja, A., et al. The interplanetary magnetic
	field from 1 to 5 AU - ULYSSES observations. \asr\vol{13}{(6) 15--24}
	1993.
\reference                                                       
Balogh, A., Gonz\'alez-Esparza, J. A., Forsyth, R. J., et al. Interplanetary
	Shock Waves: ULYSSES Observations In and Out of the Ecliptic Plane.
	\ssr\vol{72}{171--180} 1995. 
\reference                                                          
Baring, M. G., Ellison, D. C. \& Jones, F. C. Monte Carlo simulations of 
	particle acceleration at oblique shocks. \apjsupp\vol{90}{547--552} 1994.
\reference
Baring, M. G., Ogilvie, K. W., Ellison, D. C., \& Forsyth, R. J. Acceleration of 
	Solar Wind Ions by Nearby Interplanetary Shocks: Comparison of 
	Monte Carlo Simulations with Ulysses Observations.
	\apj\vol{476}{889--902} 1997.
\reference
Bell, A. R. The acceleration of cosmic rays in shock fronts. I.
	\mnras\vol{182}{147--156} 1978.
\reference
Bieber, J.~W., Matthaeus, W.~H., Smith, C.~W., et al. Proton and electron 
	mean free paths: The Palmer consensus revisited. 
	\apj\vol{420}{294--306} 1994.
\reference
Burgess, D. Alpha particles in field-aligned beams upstream of the bow 
	shock - Simulations. \grl\vol{16}{163--166} 1989.
\reference
Burton, M. E., Smith, E. J., Goldstein, B. E., et al. 
	ULYSSES - Interplanetary shocks between 1 and 4 AU.
	\grl\vol{19}{1,287--1,289} 1992.
\reference
Decker, R. B., Pesses, M. E., \& Krimigis, S. M. Shock-associated 
	low-energy ion enhancements observed by Voyagers 1 and 2.
	\jgr\vol{86}{8,819--8,831} 1981.
\reference
Ellison, D. C., Baring, M. G., \& Jones, F. C. Acceleration 
	Rates and Injection Efficiencies in Oblique Shocks.
	\apj\vol{453}{873--882} 1995.
\reference
Ellison, D. C., Baring, M. G. \&  Jones, F. C. Nonlinear 
	Particle Acceleration in Oblique Shocks.
	\apj\vol{473}{1,029--1,050} 1996.
\reference
Ellison, D. C., Giacalone, J., Burgess, D., \& Schwartz, S. J.
	Simulations of particle acceleration in parallel shocks: 
	Direct comparison between Monte Carlo and one-dimensional 
	hybrid codes. \jgr\vol{98}{21,085--21,093} 1993.
\reference
Ellison, D. C., Jones, F. C. \& Baring, M. G. Direct 
	Acceleration of Pickup Ions at the Solar Wind Termination
	Shock: The Production of Anomalous Cosmic Rays.
	\apj\vol{512}{403--416} 1999.
\reference
Ellison, D. C., Jones, F. C. \& Eichler, D. Monte Carlo 
	simulation of collisionless shocks showing preferential 
	acceleration of high A/Z particles. J. Geophys. - 
	Zeitschrift fuer Geophysik 50, 110--113. 1981.
\reference
Ellison, D. C., Jones, F. C. \& Reynolds, S. P. First-order Fermi
	particle acceleration by relativistic shocks. 
	\apj\vol{360}{702--714} 1990.
\reference
Ellison, D. C., M\"obius, E., \& Paschmann, G. Particle 
	injection and acceleration at earth's bow shock - 
	Comparison of upstream and downstream events.
	\apj\vol{352}{376--394} 1990.
\reference
Forman, M.~A., Jokipii, J.~R. \& Owens, A.~J. Cosmic-Ray Streaming 
	Perpendicular to the Mean Magnetic Field.
	\apj\vol{192}{535--540} 1974.
\reference
Giacalone, J., Burgess, D., \& Schwartz, S. J. Ion acceleration at parallel 
	shocks: Self-consistent plasma simulations, in ESA, Study of the 
	Solar-Terrestrial System, 65--70. 1992.
\reference
Giacalone, J., Burgess, D., Schwartz, S. J. \& Ellison, D. C. Ion injection
	and acceleration at parallel shocks - Comparisons of self-consistent
	plasma simulations with existing theories. \apj\vol{402}{550--559} 1993.
\reference
Giacalone, J. \& Jokipii, J.~R.The Transport of Cosmic Rays across a 
	Turbulent Magnetic Field. \apj\vol{520}{204--214} 1999.
\reference
Gloeckler, G., Geiss, J., Roelof, E. C., et al. Acceleration of interstellar
	pickup ions in the disturbed solar wind observed on ULYSSES.
	\jgr\vol{99}{17,637--17,643} 1994.
\newpage
\reference
Gonz\'alez-Esparza, J. A., Balogh, A., Forsyth, R. J., et al. Interplanetary
	shock waves and large-scale structures: Ulysses' observations in and 
	out of the ecliptic plane. \jgr\vol{101}{17,057--17,072} 1996.
\reference
Gosling, J. T., Asbridge, J. R., Bame, S. J., et al. Interplanetary ions 
	during an energetic storm particle event - The distribution function 
	from solar wind thermal energies to 1.6 MeV. \jgr\vol{86}{547--554} 
	1981.
\reference
Gosling, J. T., Thomsen, M. F., Bame, S. J., \& Russell, C. T. On the 
	source of diffuse, suprathermal ions observed in the vicinity of 
	the earth's bow shock. \jgr\vol{94}{3,555--3,563} 1989.
\reference
Hoang, S., Lacombe, C., Mangeney, A., et al. Interplanetary shocks 
	observed by ULYSSES in the ecliptic plane as a function of the 
	heliocentric distance. \asr\vol{15\, (8/9)}{371--374} 1995.
\reference
Hoppe, M. M., Russell, C. T., Frank, L. A., et al. Upstream 
	hydromagnetic waves and their association with backstreaming 
	ion populations - ISEE 1 and 2 observations.
	\jgr\vol{86}{4,471--4,492} 1981.
\reference
Jokipii, J.~R. \& Coleman, P.~J. Cosmic-Ray Diffusion Tensor 
	and Its Variation Observed with Mariner 4. \jgr\vol{73}{5,495--5,501} 1968.
\reference
Jones, F.~C., Birmingham, T.~J., \& Kaiser, T.~B. Partially averaged 
	field approach to cosmic ray diffusion. Phys. Fluids
	\vol{21}{347--360} 1978.
\reference
Jones, F. C. and Ellison, D. C. The plasma physics of shock 
	acceleration. \ssr\vol{58}{259--346} 1991.
\reference
Kang, H., \& Jones, T. W. Diffusive Shock Acceleration 
	Simulations: Comparison with Particle Methods and Bow 
	Shock Measurements. \apj\vol{447}{944--961} 1995.
\reference
Kang, H., \& Jones, T. W. Diffusive Shock Acceleration in 
	Oblique Magnetohydrodynamic Shock: Comparison with 
	Monte Carlo Methods and Observations. 
	\apj\vol{476}{875--888} 1997.
\reference
Kucharek, H. \& Scholer, M. Injection and acceleration of
	interstellar pickup ions at the heliospheric 
	termination shock. \jgr\vol{100}{1,745--1,754} 1995.
\reference
Kulsrud, R. \& Pierce, W.~P. The Effect of Wave-Particle Interactions 
	on the Propagation of Cosmic Rays. \apj\vol{156}{445--469} 1969.
\reference
le Roux, J. A., Potgieter, M. S., \& Ptuskin, V. S. A 
	transport model for the diffusive shock acceleration and
	modulation of anomalous cosmic rays in the heliosphere.
	\jgr\vol{101}{4,791--4,804} 1996.
\reference
Liewer, P. C., Goldstein, B. E., \& Omidi, N. Hybrid simulations
	of the effects of interstellar pickup hydrogen on the solar
	wind termination shock. \jgr\vol{98}{15,211--15,220} 1993.
\reference
Mason, G. M., Gloeckler, G., \& Hovestadt, D. Temporal variations of 
	nucleonic abundances in solar flare energetic particle events. 
	I - Well-connected events. \apj\vol{267}{844--862} 1983.
\reference
M\"obius, E., Scholer, M., Sckopke, N., et al. The distribution 
	function of diffuse ions and the magnetic field power spectrum 
	upstream of earth's bow shock. \grl\vol{14}{681--684} 1987.
\newpage
\reference
Quest, K. B. Theory and simulation of collisionles parallel shocks.
	\jgr\vol{93}{9,649--9,680} 1988.
\reference
Sarris, E. T., \& Van Allen, J. A. Effects of interplanetary shock 
	waves on energetic charged particles.
	\jgr\vol{79}{4,157--4,173} 1974.
\reference
Scholer, M., Hovestadt, D., Ipavich, F. M., \& Gloeckler, G. 
	Conditions for acceleration of energetic ions greater than 30 
	keV associated with the earth's bow shock. 
	\jgr\vol{85}{4,602--4,606} 1980.
\reference
Tan, L. C., Mason, G. M., Gloeckler, G., \& Ipavich, F. M. 
	Downstream energetic proton and alpha particles during 
	quasi-parallel interplanetary shock events.
	\jgr\vol{93}{7,225--7,243} 1988.
\reference
Trattner, K. J., \& Scholer, M. Diffuse alpha particles upstream 
	of simulated quasi-parallel supercritical collisionless 
	shocks. \grl\vol{18}{1,817--1,820} 1991.
\reference
Vasyliunas, V. M., \& Siscoe, G. L. On the flux and the energy 
	spectrum of interstellar ions in the solar system. 
	\jgr\vol{81}{1,247--1,252} 1976.
\reference
Winske, D., Thomas, V. A., Omidi, N., \& Quest, K. B. Re-forming
	supercritical quasi-parallel shocks. II - Mechanism for 
	wave generation and front re-formation.
	\jgr\vol{95}{18,821--18,832} 1990.

\end{document}